# Refining Wi-Fi Based Indoor Localization with Li-Fi Assisted Model Calibration in Smart Buildings

Qian Huang[1], Yuanzhi Zhang[2], Zhenhao Ge[3], Chao Lu[4]

1) Assistant Prof., School of Architecture, Southern Illinois University Carbondale, Carbondale, IL, USA. Email: qhuang@siu.edu
2) Ph.D., Department of Electrical and Computer Engineering, Southern Illinois University Carbondale, Carbondale, IL, USA. Email:yzzhang@siu.edu
3) Ph.D. Department of Electrical and Computer Engineering, Purdue University, West Lafayette, IN, USA. Email:zhenhao.ge@gmail.com
4) Assistant Prof., Department of Electrical and Computer Engineering, Southern Illinois University Carbondale, Carbondale, IL, USA. Email:chaolu@siu.edu

**Abstract:**

In recent years, there has been an increasing number of information technologies utilized in buildings to advance the idea of "smart buildings". Among various potential techniques, the use of Wi-Fi based indoor positioning allows to locate and track smartphone users inside a building, therefore, location-aware intelligent solutions can be applied to control and of building operations. These location-aware indoor services (*e.g.*, path finding, internet of things, location based advertising) demand real-time accurate indoor localization, which is a key issue to guarantee high quality of service in smart buildings. This paper presents a new Wi-Fi based indoor localization technique that achieves significantly improvement of indoor positioning accuracy with the help of Li-Fi assisted coefficient calibration. The proposed technique leverages indoor existing Li-Fi lighting and Wi-Fi infrastructure, and results in a cost-effective and user-convenient indoor accurate localization framework. In this work, experimental study and measurements are conducted to verify the performance of the proposed idea. The results substantiate the concept of refining Wi-Fi based indoor localization with Li-Fi assisted computation calibration.

## 1. INTRODUCTION

Building is an art of architecture design, heating, ventilating and air conditioning (HVAC) facility and information technology. Emerging buildings are supposed to integrate various functions to residents and provide enhanced quality of service. The rapid advance of information technologies, such as wireless sensor network, internet of things, big data and smartphones, results in smart buildings.

Smart building should adapt to accommodate their residents. In order to improve resident comfort and user experience, it is very desirable for smart buildings to know the locations of each occupant and then provide location based services, such as intelligent car parking, heath monitoring, navigation, logistics and shopping assistance (Greene, 2014). For example, in a smart shopping mall, internet of things technology connects every item for sale through wireless internet. Various retailers have implemented smartphone/tablet based indoor localization technology, whose computation algorithm is based on received signal strength indication (RSSI) of Wi-Fi signals at shopper's smartphones/tablets (Chon, 2011) (Alzantot, 2012) (Mahamud, 2015) (Bahl, 2015). The benefits of location based shopping assistance include knowing shopper's location and trajectory, conducting shopping history data analysis, learning user's shopping interest and preference, and then offering product recommendation and advertisement. Under such a circumstance, the effectiveness of shopping assistance heavily depends on the estimation accuracy of indoor localization technology. Nowadays, several companies (*e.g.,* IndoorAtlas, Skyhook Wireless, Google, and Apple) are developing their sophisticated Wi-Fi based indoor positioning applications for smartphones/tablets. They are serving over hundreds of millions of users now.

In addition to indoor shopping assistance, location based service also plays a crucial role in reduction of building energy cost. According to the U.S. Green Building Council, building contributes a significant portion of energy consumption in the United States. It is reported that buildings account for 70% of electricity load and 39% of carbon oxide emissions. Much of these energy usage could be eliminated through energy-efficient HVAC systems (Huang, 2010). In the past decades, mechanical engineers have been striving to boost energy efficiency of HVAC facilities, so users can save substantial building energy bill. Given an occupant's accurate indoor position, a building management system (BMS) is capable to offer highly efficient heating, cooling, ventilation and lighting service to the occupant. For example, when an occupant watches TV in a living room, a building management system is aware of his/her accurate location, then, it may turn down the operation of ventilation fans or air valves in other rooms of this building to reduce energy consumption. When the occupant moves to another room (*i.e.*, different thermal zone), the building management system will keep tracking her/his location and adaptively adjust

HVAC operation status. From this example, it is apparent that accurate indoor localization is of great importance for location based HVAC facility operation. As people may spend more than 80% of their time indoors, a smart building with the capability of indoor location based services has great potential to largely enhance the building resident experience and reduce building energy consumption.

Up to date, a variety of Wi-Fi based indoor positioning approaches have been proposed and developed (Kitashuka, 2003) (Youssef, 2005) (Chon, 2011) (Alzantot, 2012). Even though fundamental propagation mechanism of Wi-Fi signal has been well investigated by wireless communication researchers, yet, due to the complicated signal reflection, multipath and interference with indoor obstacles (*e.g.*, walls, windows, furniture), the received Wi-Fi signal strength at user's smartphone/tablet demonstrates a large variation range (Henniges, 2012). As a result, the indoor position prediction based on Wi-Fi's RSSI is not accurate (usually with an estimation error of several meters) and high quality of location based services is hard to obtain (Henniges, 2012) (Yim, 2013) (Bahl, 2015). This drawback limits the wide deployment of these location based services, such as shopping assistance or energy-efficient HVAC operation.

In this paper, the feasibility of indoor hybrid Wi-Fi and Li-Fi based positioning technique is investigated. Experimental tests have been conducted to illustrate the large variation of RSSI values at different indoor locations. To deal with the estimate accuracy challenge, a Wi-Fi positioning algorithm with Li-Fi assisted coefficient calibration is proposed. The integration of Li-Fi lighting infrastructure into indoor Wi-Fi based positioning increases the location estimation accuracy by 80% without any system implementation overhead. Thus, the proposed approach is a performance-refined and cost-effective solution in emerging smart buildings.

## 2. RELATED WORK

### 2.1 Overview of Li-Fi Technology

Li-Fi stands for Light-Fidelity, which refers to "visible light communication (VLC)" using light-emitting diodes (LED) light as wireless data transmission medium. The basic idea of Li-Fi technology is to utilize the visible light from an LED light bulb to transmit high speed data to a photo detector, which is connected to a smartphone or tablet. The photo detector converts the received lighting signal into electrical signal, which a smartphone or tablet can recognize and proceed easily. In fact, both Wi-Fi and Li-Fi are electromagnetic waves: Wi-Fi's frequency spectrum is around 2.4GHz or 5GHz, while Li-Fi's frequency spectrum is located in visible light band. Considering the wide-spread use of LED blubs inside buildings and large bandwidth of visible light, Li-Fi technology is much cheaper and more environmentally friendly than Wi-Fi. Li-Fi technique has great potential in many popular applications, such as location based services, mobile connectivity, smart lighting, and hazardous environments. The technical details and overview of Li-Fi could be found at http://purelifi.com/.

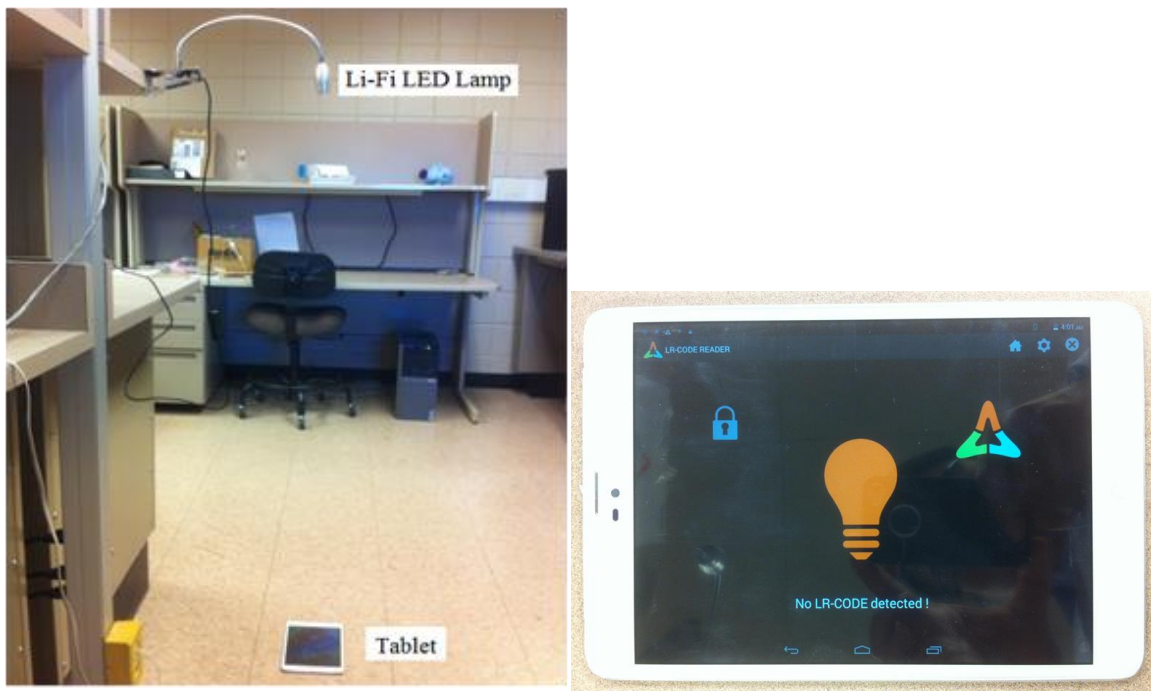

(a)          (b)
Figure 1. (a) Li-Fi communication setup (b) Tablet screenshot when no Li-Fi signal received

In this study, a GEO-LiFi test kit is chosen to carry out Li-Fi communication experiments (Oledcomm 2015). This test kit consists of several Li-Fi LED lamps, a tablet with embedded photo detector, and a location based software (*i.e.,* LR-Code Reader). If the photo detector in the tablet receives the Li-Fi signal from the Li-Fi LED lamps, the software will be triggered and execute an action (*e.g.,* pop up window, play a video).

Figure 1(a) shows an experimental setup of a Li-Fi based communication, where the LED lamp sends out the data to the tablet. If this Li-Fi signal is not detected by the embedded photo detector, the tablet window will be shown as in Figure 1(b). When the Li-Fi communication is detected, this tablet will run an action. In this test, as shown in Figure 2, the pre-loaded building floor plan will pop up and the exact location of this Li-Fi lamp will be highlighted in the floor plan.

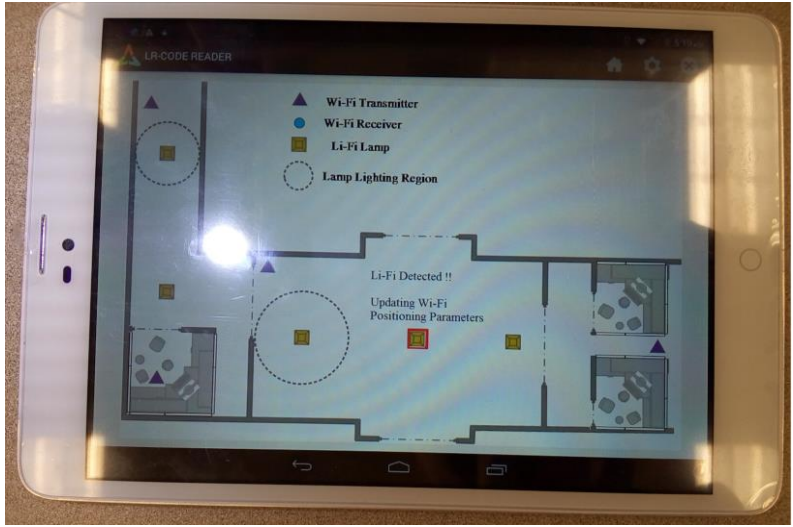

Figure 2. Smartphone screenshot when Li-Fi signal is detected

The increasing growth in the use of LED lamps in buildings for lighting provides great opportunities for Li-Fi based applications. As Li-Fi combines the functions of indoor lighting infrastructure and high-speed wireless data communication, it is very cost-effective to be widely utilized in smart buildings. In this sense, as long as the presence of a nearby LED lamp, smart building residents can get access to internet service easily without extra hardware cost (*e.g.,* Wi-Fi router or cables). When smart building residents walk inside a smart building, the indoor LED lighting will offer free wireless internet connection service. Since the location of LED lamps inside a building is usually fixed and not changing with time, Li-Fi signals received from these LED lamps can be treated as a very accurate landmark or benchmark of indoor localization. The design considerations and details of a hybrid indoor localization algorithm will be discussed in Section 3.

**2.2 Overview of Indoor Wi-Fi Based Localization Mechanism**

Indoor Wi-Fi based positioning is analogous to outdoor GPS. In an indoor Wi-Fi based positioning system, the RF received signal strength indictors (*i.e., $P_X$*) are substituted into a propagation model to calculate the distance $d$ between the Wi-Fi receiver and transmitter. This specific propagation model is expressed as (Liu, 2012)

$$P_X = P_0 - 10n\log\left(\frac{d}{d_0}\right) \tag{1}$$

where  $P_X$: received power at node $X$,
$P_0$: received power at reference node,
$d$: distance between node $X$ and Wi-Fi transmitter,
$d_0$: distance between reference node and Wi-Fi transmitter
$n$: a parameter that models the behavior of the environment (Arias, 2004) (Liu, 2012)

Alternatively, the distance $d$ and parameter $n$ can be expressed as

$$d = d_0 \times \exp\left(\frac{P_0 - P_X}{10n}\right) \tag{2}$$

$$n = \frac{P_0 - P_X}{10\log\left(\frac{d}{d_0}\right)} \tag{3}$$

Trilateration technique, where three or more Wi-Fi transmitters are required, is the primary computation approach to determine the location of a user (Liu, 2012) (Shchekotov, 2014). To improve the localization accuracy of trilateration technique, researchers have proposed a fingerprinting based Wi-Fi RSSI localization approach (Yim, 2008) (Shin, 2010), which consists of an off-line phase and an online phase. During the off-line phase, a look-up table is built to store the measured RSSI value for each location of this building. During the online phase, this look-up table will be used to match real-time measured RSSI values and determine the user location. The disadvantage is that the fingerprint database needs to be updated once the building environment (*e.g.,* furniture adding or removing) changes, therefore, it involves labor-intensive and user-unfriendly process to maintain fingerprint database up to date. It is very attractive to create a novel Wi-Fi based localization approach, which can achieve refined accuracy with little overhead of maintaining updated database. This is focus of this paper.

In this work, eZ430-RF2500-SHE system kit from Texas Instruments is used to measure RSSI values of Wi-Fi access points. Figure 3 shows the pictures of Wi-Fi receiver and transmitter, as well as the monitoring software snapshots. The Wi-Fi signal is located at 2.4 GHz. Table 1 lists the Wi-Fi signal strengths of five typical measurements at different locations. This table illustrates that for the same distance between Wi-Fi access point and Wi-Fi receiver, the received signal strength has a remarkable variations. The derived parameter *n* changes from 0.2337 to 0.4006 based on the location of measurements. As plotted in Figure 3, there are significant estimation errors if the theoretical equation neglects the impact of indoor environments by assuming a constant *n*.

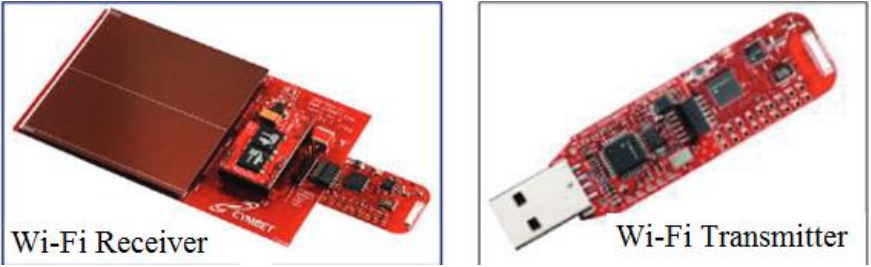

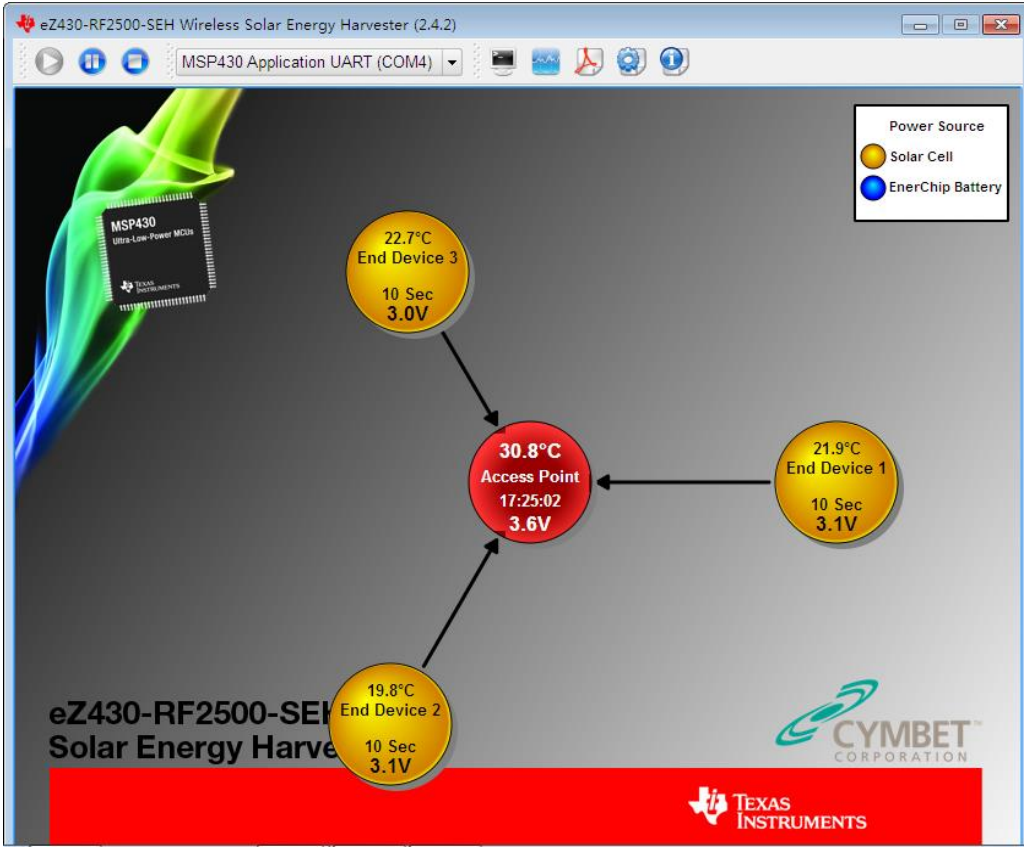

Figure 3. Wi-Fi communication hardware and software snapshot: three Wi-Fi access points (yellow icons) send signals to Wi-Fi receiver (red icon)

Table 1. Normalized RSSI Values at Different Indoor Locations

| Distance | Location #1 | Location #2 | Location #3 | Location #4 | Location #5 |
|---|---|---|---|---|---|
| 1 meter | 26% | 26% | 24% | 27% | 25% |
| 5 meter | 20% | 23% | 19% | 21% | 23% |
| 10 meter | 19% | 22% | 17% | 18% | 21% |
| 15 meter | 17% | 21% | 16% | 16% | 18% |
| 20 meter | 16% | 19% | 15% | 15% | 16% |
| Derived $n$ value in equation (1) | 0.3338 | 0.2337 | 0.3004 | 0.4006 | 0.3004 |

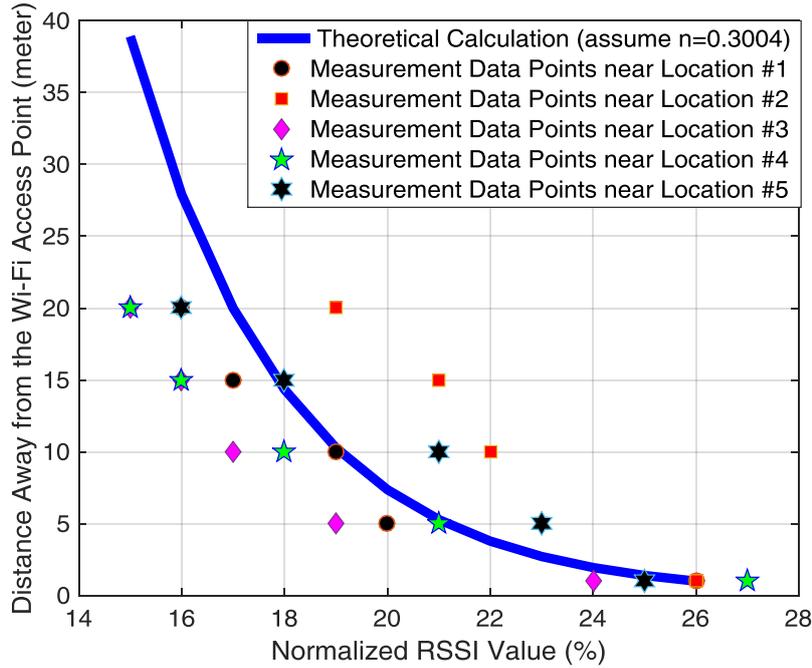

Figure 4. Theoretical calculation results vs. measurement data points

## 3. INDOOR WI-FI POSITIONING ALGORITHM WITH LI-FI ASSISTED CALIBRATION

Figure 5 represents the ground-level floor plan of electrical and computer engineering building of Southern Illinois University Carbondale during the testing. This level consists of classrooms, atrium, labs and aisles. Positions of Wi-Fi transmitters and Li-Fi LED lamps are marked in this figure. The Wi-Fi receiver keeps moving and receives the RSSI from these Wi-Fi transmitters.

Figure 6 exhibits the proposed flowchart of indoor Wi-Fi localization algorithm with Li-Fi assisted coefficient calibration. Once a person enters a building, the indoor localization function will start if he/she wants. The Li-Fi software "LR-CODE READER" is activated to detect surrounding Li-Fi signals. Meanwhile, the Wi-Fi signals are received by Wi-Fi antenna of this smartphone or tablet. Once the Li-Fi photo detector is successfully triggered, the smartphone/tablet knows the identifier of this specific Li-Fi LED lamp. The current location of this person is indicated within the Li-Fi lamp lighting region as depicted in Figure 5. Thus, according to the building floor map, the distance between Wi-Fi access point and this identified Li-Fi lamp is obtained. Substituting this distance back to the theoretical equation 3 in Section 2.2 leads to a derived coefficient value (*i.e.*, parameter *n* in equation 3). Since moving speed of a person inside buildings is not very fast, the derived coefficients will be used in real-time Wi-Fi based localization until next triggering time of Li-Fi photo detector at different LED lamp regions. Based on any RSSI values and calibrated coefficient value, the smartphone or tablet computes the indoor location according to trilateration principle.

The advantage of this proposed hybrid method eliminates the necessity of frequent update of look-up tables in the fingerprinting approach. There is no look-up table required in this proposed approach. The computation coefficient values are derived immediately after the triggering of Li-Fi photo detector. As a result, no matter what kind of change is applied in the indoor building environment (e.g., adding or removing furniture), this derived parameter value reflects this environmental change.

From the above discussion, we can see that when a person keeps moving inside a building, the Li-Fi photo detector will be triggered continuously. After each triggering event, the coefficient values for computing indoor localization is updated automatically. As a result, this proposed hybrid Wi-Fi and Li-Fi localization technique has a great potential to achieve performance-refined and user-friendly positioning.

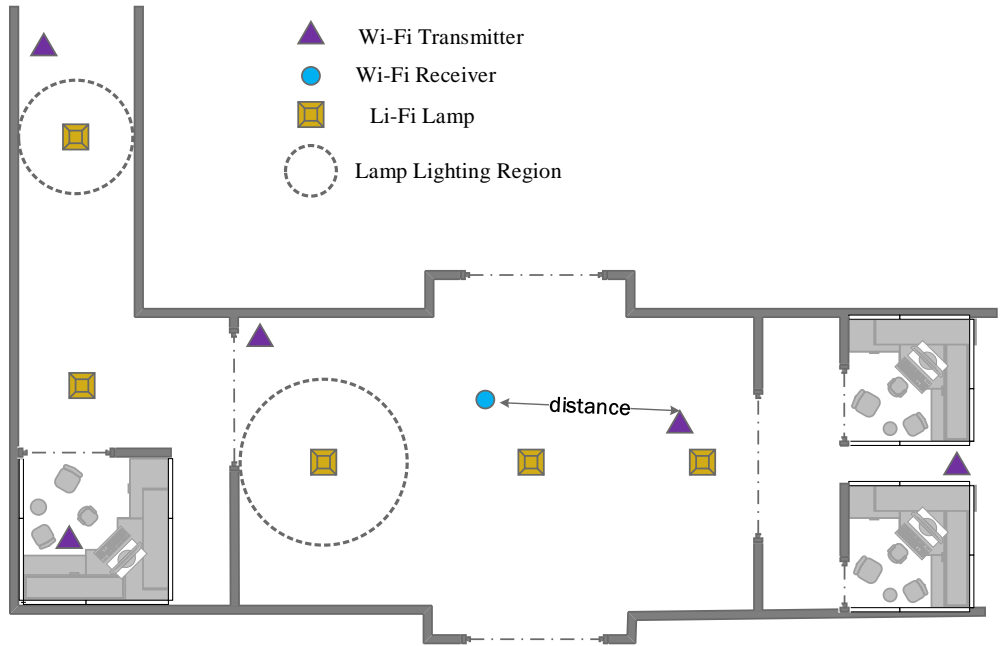

Figure 5. Building floor plan with Wi-Fi access points and Li-Fi LED lamps

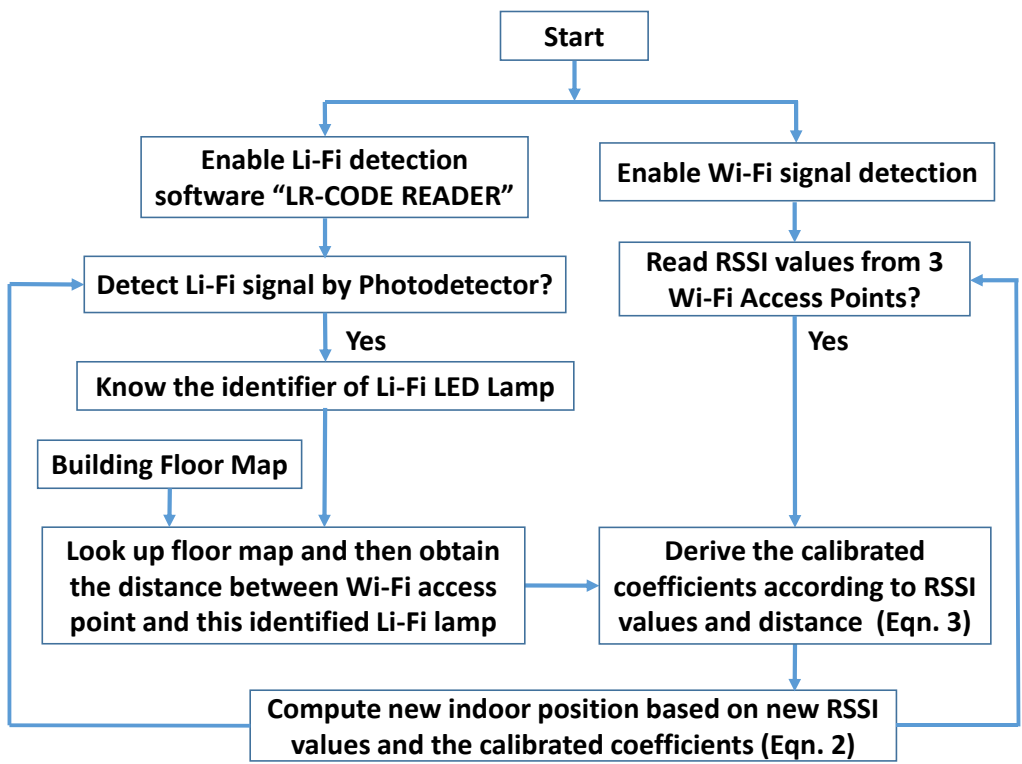

Figure 6. Flowchart of indoor Wi-Fi positioning algorithm with Li-Fi assisted calibration

Figure 7 shows the localization performance of our proposed algorithm. When a person walks into location #1, the Li-Fi photo detector helps to update and derive a computation coefficient n=0.3338. Thus, the theoretical model will be moved from the solid line to the dashed line as plotted in Figure 7(a). In Figure 7(a), the averaged prediction error between the solid line and these five measurement data points are 3.054 meters, while the averaged prediction error between the dashed line and these five measurement data points are 0.618 meters. This example accounts for an 80% reduction of estimation error. Figure 7(b) plots an even better performance when the proposed hybrid algorithm is applied for location #4.

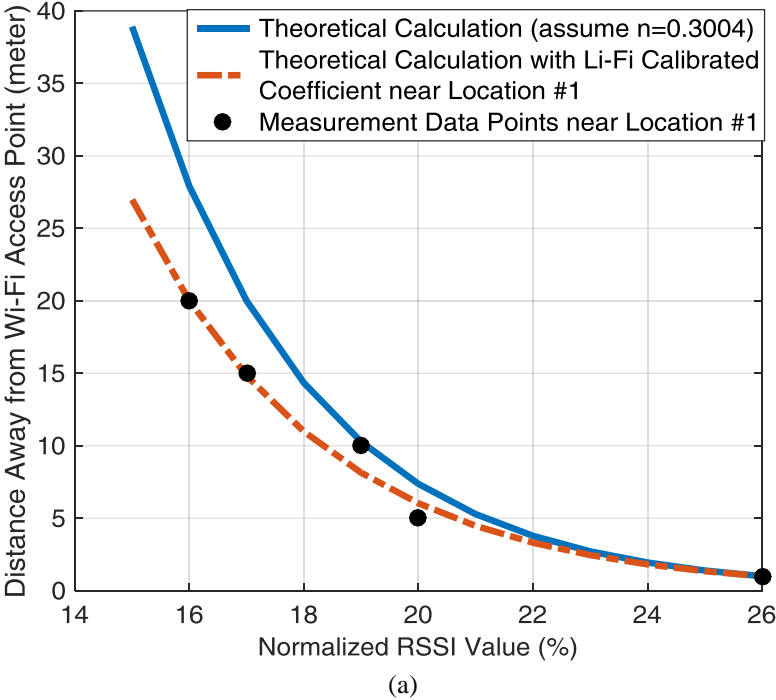

(a)

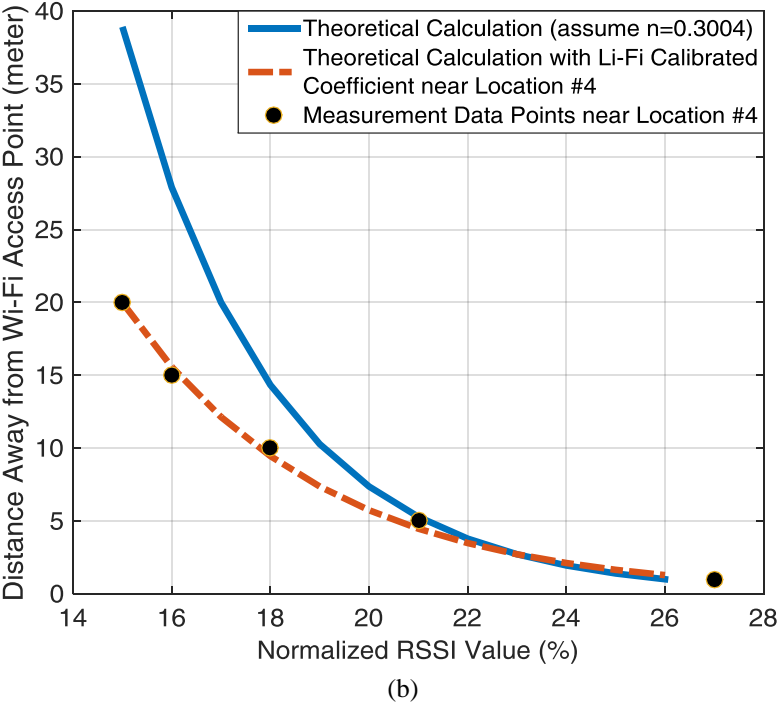

(b)

Figure 7. Indoor hybrid Wi-Fi and Li-Fi localization results vs. conventional Wi-Fi localization results vs. measurement data points

## 4. CONCLUSIONS

Location-aware indoor service in smart buildings demand accurate indoor localization technique. The proposed indoor hybrid Li-Fi and Wi-Fi positioning system can achieve more precise location estimation than existing Wi-Fi based positioning systems. Experimental results in this work demonstrate an accuracy improvement of 80%.

In contrast with fingerprinting based Wi-Fi RSSI localization technique, this proposed approach does not require the frequent update of fingerprinting look-up table. Therefore, the proposed approach does not involve labor-intensive and user-unfriendly process to maintain fingerprint database up to date. By leveraging indoor Li-Fi lighting infrastructure, this proposed scheme does not increase the system implementation cost. This proposed scheme is promising to overcome the large variation of estimation in Wi-Fi based positioning systems in smart buildings, where Li-Fi LED lamps have been widely installed as Li-Fi access points.